\documentclass[aps,nofootinbib,prd,superscriptaddress,tightenlines,notitlepage,twocolumn,showpacs]{revtex4}

\usepackage[colorlinks]{hyperref}
\usepackage{tabularx}
\usepackage{graphicx}
\usepackage{amssymb,bm}
\usepackage{xcolor}
\usepackage{amsmath}
\usepackage[varg]{txfonts}
\usepackage{enumerate}
\usepackage[normalem]{ulem}
\usepackage{subfigure}

\newcommand{\PMO}{\affiliation{Purple Mountain Observatory, Chinese Academy of Sciences, Nanjing 210023, China}}
\newcommand{\USTC}{\affiliation{School of Astronomy and Space Sciences, University of Science and Technology of China, Hefei 230026, China}}

\begin{document}

\title{Extragalactic Test of General Relativity from Strong Gravitational Lensing by using Artificial Neural Networks}

\author{Jing-Yu Ran}\PMO\USTC
\author{Jun-Jie Wei}\thanks{jjwei@pmo.ac.cn}\PMO\USTC

\begin{abstract}
This study aims to test the validity of general relativity (GR) on kiloparsec scales
by employing a newly compiled galaxy-scale strong gravitational lensing (SGL) sample.
We utilize the distance sum rule within the Friedmann-Lema\^{\i}tre-Robertson-Walker
metric to obtain cosmology-independent constraints on both the parameterized post-Newtonian parameter
$\gamma_{\rm PPN}$ and the spatial curvature $\Omega_{k}$, which overcomes the
circularity problem induced by the presumption of a cosmological model grounded in
GR. To calibrate the distances in the SGL systems, we introduce a novel nonparametric
approach, Artificial Neural Network (ANN), to reconstruct a smooth distance--redshift
relation from the Pantheon+ sample of type Ia supernovae. Our results show that
$\gamma_{\rm PPN}=1.16_{-0.12}^{+0.15}$ and $\Omega_k=0.89_{-1.00}^{+1.97}$, indicating
a spatially flat universe with the conservation of GR (i.e., $\Omega_k=0$ and $\gamma_{\rm PPN}=1$)
is basically supported within $1\sigma$ confidence level. Assuming
a zero spatial curvature, we find $\gamma_{\rm PPN}=1.09_{-0.10}^{+0.11}$,
representing an agreement with the prediction of 1 from GR to a 9.6\% precision.
If we instead assume GR holds (i.e., $\gamma_{\rm PPN}=1$), the curvature parameter
constraint can be further improved to be $\Omega_k=0.11_{-0.47}^{+0.78}$.
These resulting constraints demonstrate the effectiveness of our method in testing
GR on galactic scales by combining observations of strong lensing and the
distance--redshift relation reconstructed by ANN.
\end{abstract}



\maketitle

\section{Introduction}\label{sec:intro}
As an important cornerstone of modern physics, Einstein's theory of general relativity
(GR) has withstood very strict tests (e.g., \citep{1920RSPTA.220..291D,
1960PhRvL...4..337P,PhysRevLett.13.789,1979Natur.277..437T}). But testing GR at a much
higher precision is still a vital task, because any possible violation of GR would
have profound effects on our understanding of fundamental physics. Within the parameterized
post-Newtonian (PPN) formalism, GR predicts that the PPN parameter $\gamma_{\rm PPN}$
which describes the amount of space curvature produced per unit rest mass should be exactly
1 \cite{1971ApJ...163..595T}. Measuring $\gamma_{\rm PPN}$ therefore serves as a test of
the validity of GR on large scales. That is, any deviation from $\gamma_{\rm PPN}=1$
implies a possible violation of GR.

On solar system scales, the GR prediction for $\gamma_{\rm PPN}$ has been confirmed with
high accuracy. By measuring the round-trip travel time of radar signals passing near the
Sun, the Cassini spacecraft yielded $\gamma_{\rm PPN}=1+(2.1\pm 2.3)\times 10^{-5}$
\citep{2003Natur.425..374B}. However, the extragalactic tests of GR are still insufficiency
and much less precise. On galactic scales, strong gravitational lensing (SGL), combined with
stellar kinematics in the lensing galaxy, provides an effective way to test the validity of GR
by constraining the PPN parameter $\gamma_{\rm PPN}$. The pioneering work by Ref.~\cite{2006PhRvD..74f1501B}
first utilized this approach and reported a result of $\gamma_{\rm PPN}=0.98\pm 0.07$ based on
observations of 15 elliptical lensing galaxies from the Sloan Lens ACS (SLACS) Survey.
Since then, numerous studies have been conducted to test GR using different SGL samples \citep{Smith.2009arXiv0907.4829S,2010ApJ...708..750S,2017ApJ...835...92C,Collett2018Sci...360.1342C,
Yang:2020eoh,2022ApJ...927...28L,2022ApJ...927L...1W}. In this paper, we further explore
the validity of GR by employing a newly compiled SGL sample \citep{10.1093/mnras/stz1902},
which consists of 161 galaxy-scale strong lensing systems. This larger SGL sample allows us
to perform a more comprehensive analysis and obtain further insights into the behavior of
gravity on galactic scales.

In practice, in order to constrain the PPN parameter $\gamma_{\rm PPN}$ using SGL systems,
one has to know a ratio of three angular diameter distances (i.e., the distances from
the observer to the lens, $D_l$, the observer to the source, $D_s$, and the lens to the source, $D_{ls}$).
In most previous works, the required distance ratio is calculated within the context of the standard
$\Lambda$CDM cosmological model. However, $\Lambda$CDM itself is established on the framework
of GR, which leads to a circularity problem in testing GR \citep{2022ApJ...927...28L,2022ApJ...927L...1W}.
To circumvent this problem, we will introduce the distance sum rule (DSR) in the
Friedmann-Lema\^{\i}tre-Robertson-Walker (FLRW) metric. The two distances $D_l$ and $D_s$
can be directly determined from observations of type Ia supernovae (SNe Ia), but not the distance
$D_{ls}$. The DSR enables us to convert $D_{ls}$ into a relationship with $D_l$, $D_s$, and
the spatial curvature $\Omega_{k}$. Based on the DSR in the FLRW metric, cosmology-independent
constraints on both $\gamma_{\rm PPN}$ and $\Omega_{k}$ can thus be obtained by combing observations
of strong lensing and SNe Ia \citep{2017ApJ...835...92C,2022ApJ...927L...1W}.

Very recently, by employing the Gaussian Process (GP) method, \citet{2022ApJ...927...28L} reconstructed
a smooth distance--redshift relation directly from SN Ia observation to calibrate the distances
in the SGL sample, however, with a grossly underestimated error. GP allows for the reconstruction
of a function from a dataset without assuming
a specific model or parameterization, and it has been widely used in cosmological researches
\citep{Seikel2012UsingHD,2014PhRvD..89b3503Y,Montiel2014PhRvD..89d3007M,Wei2017ApJ...838..160W,
Melia2018MNRAS.481.4855M,Liao2019ApJ...886L..23L,2021MNRAS.506L...1D}. In the GP analysis, the errors in the observational
data are assumed to follow a Gaussian distribution \citep{2012JCAP...06..036S}. However, the actual
observations might not follow Gaussian distributions. This may thus be a strong assumption for
reconstructing a function from observational data. Moreover, due to the sparsity and scatter of
data points at high redshifts, the GP reconstructed function from SN Ia data exhibits strange oscillations with large uncertainties. To address these concerns and ensure the reliability of
the reconstructed function, we employ the Artificial Neural Network (ANN) method, which is a machine learning technique and has been proven to be a ``universal approximator'' that can reconstruct a great variety of functions \citep{Cybenko1989ApproximationBS,HORNIK1991251}. Thanks to the powerful property of neural networks, methods based on ANNs have been widely used in regression and estimation tasks.
In this work, we will reconstruct the distance--redshift relation from SN Ia data using the ANN method, utilizing a code developed in Ref~\citep{2020ApJS..246...13W}.

This paper is organized as follows: in Section \ref{sec:method}, we introduce the methodology and
observations used for testing GR on galactic scales. Cosmology-independent constraints on
$\gamma_{\rm PPN}$ and $\Omega_{k}$ are shown in Section \ref{sec:result}. In Section \ref{sec:c&d},
we make a summary and end with some discussions.

\section{Methodology and Data} \label{sec:method}
In the weak-field limit, the metric of space-time can be characterized as
\begin{equation}
\mathrm{d}s^{2}=c^{2}\mathrm{d}t^{2}\left(1-\frac{2GM}{c^{2}r}\right)-\mathrm{d}r^{2}\left(1+\frac{2\gamma_{\rm PPN} GM}{c^{2}r}\right)-r^{2}\mathrm{d}\Omega^{2}\;,
\end{equation}
where $\gamma_{\rm PPN}$ is the PPN parameter, $M$ is the mass of the central object, and $\Omega$ is the angle
in the invariant orbital plane. In the framework of GR, $\gamma_{\rm PPN}$ is equal to unity.

\subsection{Gravitational Lensing Theory} \label{subsec:SGL}
The main idea of testing the validity of GR via SGL systems is that the mass enclosed within the Einstein radius
derived separately from the gravitational theory and the dynamical theory should be equivalent, i.e.,
$M_{\rm E}^{\rm grl}=M_{\rm E}^{\rm dyn}$. From the theory of gravitational lensing \citep{1992grle.book.....S},
the Einstein angle $\theta_{\rm E}$ reflecting the angular separations between multiple images is related to
the gravitational mass $M_{\rm E}^{\rm grl}$,
\begin{equation}\label{2}
\theta_{\rm E}=\sqrt{\frac{1+\gamma_{\rm PPN}}{2}}\left ( \frac{4GM_{\rm E}^{\rm grl}}{c^2}\frac{D_{ls}}{D_{l}D_{s}}\right )^{1/2}\;,
\end{equation}
where $D_{l}$, $D_{s}$, and $D_{ls}$ are, respectively, the angular diameter distances from the observer to the lens,
the observer to the source, and the lens to the source. By introducing the Einstein radius $R_{\rm E}=D_l \theta_{\rm E}$,
Equation~(\ref{2}) can be rearranged as
\begin{equation}\label{3}
\frac{GM_{\rm E}^{\rm grl}}{R_{\rm E}}=\frac{2}{1+\gamma_{\rm PPN}}\frac{c^2}{4}\frac{D_{s}}{D_{ls}}\theta_{\rm E}\;.
\end{equation}

To estimate the dynamical mass $M_{\rm E}^{\rm dyn}$ from the spectroscopic measurement of the lens velocity dispersion,
one must first set a mass distribution model for the lensing galaxy. Here we use the common mass model with power-law
density profiles \citep{2006ApJ...649..599K,10.1093/mnras/stz1902}:
\begin{equation}\label{4}
\left\{\begin{array}{l}
\rho(r)=\rho_{0}\left(\frac{r}{r_0}\right)^{-\alpha} \\
\nu(r)=\nu_{0}\left(\frac{r}{r_0}\right)^{-\delta} \\
\beta(r)=1-\sigma_{t}^{2} / \sigma_{r}^{2}\;,
\end{array}\right.
\end{equation}
where $r$ is defined as the spherical radial coordinate from the lens centre, $\rho\left ( r \right )$ is the total
(including luminous and dark matter) mass density distribution, and $\nu \left ( r \right )$ represents the distribution
of luminous density. The parameter $\beta\left ( r \right )$ describes the anisotropy of the stellar velocity dispersion,
where $\sigma_t$ and $\sigma_r$ are the velocity dispersions in the tangential and radial directions, respectively.
In the literature, $\beta$ is always assumed to be independent of $r$ (e.g.,
\citep{2006ApJ...649..599K,2010ApJ...709.1195T}).
Following previous studies \citep{2006PhRvD..74f1501B,2010ApJ...708..750S,2017ApJ...835...92C,
10.1093/mnras/stz1902,2022ApJ...927...28L,2022ApJ...927L...1W}, we set a Gaussian prior $\beta = 0.18 \pm 0.13$, informed by the constraint from a well-studied sample of elliptical galaxies \citep{2001AJ....121.1936G}. That is,
$\beta$ will be marginalized using a Gaussian prior of $\beta = 0.18 \pm 0.13$ over the $2\sigma$ range of
$[-0.08,\;0.44]$. Also, $\alpha$ and $\delta$ are the power-law indices of the total mass density profile and
the luminosity density profile, respectively. It has been confirmed in previous works
\citep{Sonnenfeld2013,10.1093/mnras/stz1902}
that $\alpha$ is significantly related with the lens redshift $z_l$ and the surface mass density of the lensing galaxy.
Therefore, we treat the parametrized model of $\alpha$ as \citep{10.1093/mnras/stz1902}
\begin{equation}\label{5}
    \alpha=\alpha_0+\alpha_zz_l+\alpha_s\log_{10}\tilde{\Sigma}\;,
\end{equation}
where $\alpha_0$, $\alpha_z$ and $\alpha_s$ are arbitrary constants. Here $\tilde{\Sigma}$ stands for
the normalized surface mass density, and is expressed as
$\tilde{\Sigma}=\frac{\left( \sigma_0/100\;\rm km\;s^{-1}\right )^2}{R_{\rm eff}/10\;h^{-1}\;\rm kpc}$,
where $\sigma_0$ is the observed velocity dispersion, $R_{\rm eff}$ is the lensing galaxy's half-light radius, and $h=H_0/(100\;\rm km\;s^{-1}\;Mpc^{-1})$ is the reduced Hubble constant.

Following the well-known radial Jeans equation in spherical coordinate \citep{1980MNRAS.190..873B}, the radial velocity dispersion of the luminous matter $\sigma_r$ in early-type lens galaxies takes the form
\begin{equation}\label{6}
    \sigma_r^2\left(r \right )=\frac{G\int_{r}^{\infty}\mathrm d{r}'{r}'^{2\beta-2}\nu\left(r'\right)M\left(r'\right)}{r^{2\beta}\nu\left(r\right)}\;,
\end{equation}
where $M\left(r\right)$ is the total mass included within a sphere with radius $r$,
\begin{equation}\label{7}
    M\left(r\right)=\int_0^r\mathrm dr'4\pi r'^2\rho\left(r'\right)=4\pi\frac{\rho_0}{r_0^{-\alpha}}\frac{r^{3-\alpha}}{3-\alpha}\;.
\end{equation}
The dynamical mass $M_{\rm E}^{\rm dyn}$ enclosed within a cylinder of radius equal to the Einstein radius $R_{\rm E}$
can be written as \citep{10.1093/mnras/stz1902}
\begin{equation}\label{8}
    M_{\rm E}^{\rm dyn}=2\pi^{3/2}\frac{R_{\rm E}^{3-\alpha}}{3-\alpha}\frac{\Gamma\left(\frac{\alpha-1}{2}\right)}{\Gamma\left(\frac{\alpha}{2}\right)}\frac{\rho_0}{r_0^{-\alpha}}\;,
\end{equation}
where $\Gamma(x)$ is Euler's Gamma function. By combining Equations (\ref{7}) and (\ref{8}), we get the relation
between $M\left(r\right)$ and $M_{\rm E}^{\rm dyn}$:
\begin{equation}\label{9}
    M(r)=\frac{2}{\sqrt{\pi}}\frac{1}{\lambda (\alpha)}\left(\frac{r}{R_{\rm E}}\right)^{3-\alpha}M_{\rm E}^{\rm dyn}\;,
\end{equation}
where $\lambda (\alpha)=\Gamma\left(\frac{\alpha-1}{2}\right)/\Gamma\left(\frac{\alpha}{2}\right)$.
By substituting Equations (\ref{9}) and (\ref{4}) into Equation (\ref{6}), we obtain
\begin{equation}\label{10}
    \sigma_r^2\left(r \right )=\frac{2}{\sqrt{\pi}}\frac{GM_{\rm E}^{\rm dyn}}{R_{\rm E}}\frac{1}{\xi -2\beta }\frac{1}{\lambda (\alpha)}\left(\frac{r}{R_{\rm E}}\right)^{2-\alpha}\;,
\end{equation}
where $\xi=\alpha+\delta-2$.

The actual velocity dispersion of the lensing galaxy is the component of luminosity-weighted average along
the line of sight and measured over the effective spectroscopic aperture $R_{\rm A}$, that can be expressed as
(see Ref.~\citep{10.1093/mnras/stz1902} for more details)
\begin{equation}\label{11}
    \sigma_0^2\left(\leq R_{\rm A} \right )=\frac{c^2}{2\sqrt{\pi}}\frac{2}{1+\gamma_{\rm PPN}}\frac{D_s}{D_{ls} }\theta_{\rm E} F\left ( \alpha,\;\delta,\;\beta  \right )\left(\frac{R_{\rm A}}{R_{\rm E}}\right)^{2-\alpha}\;,
\end{equation}
where
\begin{equation}\label{12}
    F\left( \alpha,\;\delta,\;\beta  \right)=\frac{3-\delta}{\left( \xi-2\beta \right)\left( 3-\xi\right)}\frac{\lambda(\xi)-\beta\lambda\left ( \xi+2 \right )}{\lambda(\alpha)\lambda(\delta)}\;.
\end{equation}
The theoretical value of the velocity dispersion inside the radius $R_{\rm eff}/2$ can then be calculated by
\citep{2006ApJ...649..599K}
\begin{equation}\label{13}
    \sigma_0^{\rm th}=\sqrt{\frac{c^2}{2\sqrt{\pi}}\frac{2}{1+\gamma_{\rm PPN}}\frac{D_s}{D_{ls}}\theta_{\rm E} F\left( \alpha,\;
    \delta,\; \beta  \right)\left(\frac{\theta_{\rm eff}}{2\theta_{\rm E}}\right)^{2-\alpha}}\;,
\end{equation}
where $\theta_{\rm eff}=R_{\rm eff}/D_l$ denotes the effective angular radius of the lensing galaxy.

Based on the spectroscopic data, one can measure the luminosity-weighted average of the line-of-sight velocity dispersion
$\sigma_{\rm ap}$ within the circular aperture with the angular radius $\theta_{\rm ap}$. In practice, $\sigma_{\rm ap}$
should be normalized to the velocity dispersion within the typical physical aperture with a radius $R_{\rm eff}/2$,
\begin{equation}\label{14}
    \sigma_0^{\rm obs}=\sigma_{\rm ap}\left[ \theta_{\rm eff}/(2\theta_{\rm ap}) \right]^{\eta}\;,
\end{equation}
where the value of the correction factor is taken as $\eta =-0.066\pm0.035$ \citep{2006MNRAS.366.1126C}.
Then, the total uncertainty of $\sigma_0^{\rm obs}$
can be obtained by
\begin{equation}\label{15}
    \left ( \Delta\sigma_0^{\rm SGL} \right )^2=\left ( \Delta\sigma_0^{\rm stat} \right )^2+\left ( \Delta\sigma_0 ^{\rm AC}\right )^2+\left ( \Delta\sigma_0^{\rm sys} \right )^2\;,
\end{equation}
where $\Delta\sigma_0^{\rm stat}$ is the statistical error propagated from the measurement error of $\sigma_{\rm ap}$, and
$\Delta\sigma_0 ^{\rm AC}$ is the aperture-correction-induced error propagated from the uncertainty of $\eta$. The systematic error
due to the extra mass contribution from the outer matters of the lensing galaxy along the line of sight, $\Delta\sigma_0^{\rm sys}$,
is taken as an uncertainty of $\sim3\%$ to the velocity dispersion \citep{2007ApJ...671.1568J}.

Once we know the ratio of the angular diameter distances $D_{s}/D_{ls}$, the constraints on the PPN parameter $\gamma_{\rm PPN}$
can be derived by comparing the observational and theoretical values of the velocity dispersions (see Equations~(\ref{13}) and (\ref{14})). Conventionally the distance ratio $D_{s}/D_{ls}$ is calculated within the standard $\Lambda$CDM cosmological model
\citep{2010ApJ...708..750S,2017ApJ...835...92C}. However, $\Lambda$CDM itself is built on the framework of GR and this
leads to a circularity problem \citep{2022ApJ...927...28L,2022ApJ...927L...1W}. To avoid this problem, we will use a cosmological-model-independent method which is based upon the sum rule of distances in the FLRW metric to constrain $\gamma_{\rm PPN}$.

\subsection{Distance Sum Rule} \label{subsec:DSR}
In a homogeneous and isotropic space, the dimensionless comoving distance
$d\left(z_l,\;z_s \right)\equiv \left(H_0/c\right)\left(1+z_s\right) D_A \left(z_l,\;z_s \right)$ can be written as
\begin{equation}
d(z_l,z_s)=\frac{1}{\sqrt{|\Omega_{k}|}}{\rm sinn}\left(\sqrt{|\Omega_{k}|}\int_{z_l}^{z_s}\frac{{\rm d}z'}{E(z')}\right)\;,
\label{eq:dls}
\end{equation}
where $\Omega_k$ denotes the spatial curvature density parameter at the present time and $E(z)=H(z)/H_0$ is the dimensionless
expansion rate. Also, ${\rm sinn}(x)$ is $\sinh(x)$ when $\Omega_k>0$, $x$ when $\Omega_k=0$, and $\sin(x)$ when $\Omega_{k}<0$.
By applying the notations $d(z)\equiv d\left(0,\;z\right)$, $d_{ls}\equiv d\left(z_l,\;z_s\right)$, $d_l\equiv d\left(0,\;z_l\right)$,
and $d_s\equiv d\left(0,\;z_s \right)$, one can derive a sum rule of distances along the null geodesics of the FLRW metric as
\citep{1993ppc..book.....P,2006ApJ...637..598B,2015PhRvL.115j1301R}
\begin{equation}\label{18}
     \frac{d_{ls}}{d_s}=\sqrt{1+\Omega_kd_l^2}-\frac{d_l}{d_s}\sqrt{1+\Omega_kd_s^2}\;.
\end{equation}
This relation provides a cosmology-independent probe to test both the spatial curvature and the FLRW metric.
The validity of the FLRW metric can be tested by comparing the derived $\Omega_k$ from the three distances
($d_l$, $d_s$, and $d_{ls}$) for any two pairs of ($z_l$, $z_s$).

With Equation~(\ref{18}), the distance ratio $D_{s}/D_{ls}$\footnote{Note that $D_{s}/D_{ls}$ is actually equal to  the dimensionless distance ratio $d_{s}/d_{ls}$.} in Equation~(\ref{13}) is only related to the curvature parameter
$\Omega_k$ and the dimensionless distances $d_l$ and $d_s$. If independent measurements of $d_l$ and $d_s$ are given,
we can put constraints on both $\gamma_{\rm PPN}$ and $\Omega_k$ from Equations~(\ref{13}) and (\ref{18}) without assuming any specific cosmological model.

\subsection{Artificial Neural Network} \label{subsec:ANN}
To calibrate the distances $d_l$ and $d_s$ of the SGL systems (i.e., the distances $d_l$ and $d_s$ on the right side of Equation~(\ref{18})), we use a new nonparametric approach, ANN, to reconstruct a smooth distance--redshift relation $d(z)$ from SN Ia observation.

ANNs possess several desirable properties, including high-level abstraction of neural input-output transformation, the ability to generalize from learned instances to new unseen data, adaptability, self-learning, fault tolerance, and nonlinearity \citep{ABIODUN2018e00938}. According to the universal approximation theorem \citep{HORNIK1989359, Cybenko1989ApproximationBS}, ANNs can function as universal function approximators to simulate arbitrary input-output relationships using multilayer feedforward networks with a sufficient number of hidden units. Therefore, we can input the redshift $z$ into the neural network, with the corresponding comoving distance $d(z)$ and its associated error $\sigma_{d(z)}$ as the desired outputs. Once the network has been trained using the Pantheon+ sample, we will obtain an approximate function capable of predicting both $d(z)$ and its error $\sigma_{d(z)}$ at any given redshift $z$.

Ref.~\cite{2020ApJS..246...13W} has developed a Python code for the reconstruction of functions from observational data employing an ANN. They have substantiated the reliability of these reconstructed functions by estimating cosmological parameters through the utilization of the reconstructed Hubble parameter $H(z)$ and the luminosity distance $D_L(z)$, in direct comparison with observational data. In our study, we will employ this code to reconstruct the distance--redshift relation.

The general structure of an ANN consists of an input layer, one or more hidden layers, and an output layer. The basic unit of these layers are referred to as neurons, which serve as both linear transformation units and nonlinear activation functions for the input vector. In accordance with Ref.~\cite{2020ApJS..246...13W}, we employ the Exponential Linear Unit as our chosen activation function, as defined by its form in \citep{2015arXiv151107289C}:
\begin{equation}\label{elu}
f\left ( x \right )=\left\{\begin{matrix}
x & x> 0\\
\alpha \left ( e^{x}-1 \right ) & x\leq  0
\end{matrix}\right.\;,
\end{equation}
where the hyperparameter $\alpha$ is set to 1.

The network is trained by minimizing a loss function, which quantitatively measures the discrepancy between the ground truth and predicted values. In this analysis, we adopt the mean absolute error (MAE), also known as the L1 loss function, as our choice of loss function. The linear weights and biases within the network are optimized using the backpropagation algorithm. We employ the Adam optimizer \citep{2014arXiv1412.6980K}, a gradient-based optimization technique, to iteratively update the network parameters during training. This choice of optimizer also contributes to faster convergence. After multiple iterations, the network parameters are adjusted to minimize the loss. We have determined that a sufficient number of iterations for training convergence is reached when the loss no longer decreases, which we set to be $3\times 10^{5}$. Batch Normalization \citep{2015arXiv150203167I} is a technique designed to stabilize the distribution of inputs within each layer, allowing for higher learning rates and reduced sensitivity to initialization.

To determine the optimal network model, we train the network using 1701 SNe Ia from the Pantheon+ sample (more on this below) and assess the fitting effect through K-fold cross-validation \citep{scikit-learn}. In K-fold cross-validation, the training set is divided into k smaller sets, with k-1 folds used as training data for model training, and the remaining fold used for validation. This process is repeated k times, with each fold serving as the validation data once. The final performance of the model is determined by averaging the performance across these k iterations. This approach is particularly useful when the number of samples available for learning is insufficient to split into traditional train, validation, and test sets, as is the case in our analysis. Additionally, it helps mitigate issues arising from the randomness in data partitioning. As a general guideline, we have selected $k=10$ for our cross-validation procedure and have utilized the mean squared error (MSE) as the metric for validating the performance of the model.

Through our experimentation, we have found that the network model with a single hidden layer comprising 4096 neurons and without batch normalization yields the best results. We conduct comparisons with models having varying numbers of hidden layers, and we observe diminished performance as the number of hidden layers increased, accompanied by increased computational resource consumption. Regarding the number of neurons in the hidden layer, we observe negligible impact on the results, as reflected by the final MSE values consistently hovering around 0.0042, regardless of whether the number of neurons was set to 1024, 2048, 4096, or 8192. Importantly, the final MSE value with 4096 neurons was slightly smaller compared to the other three configurations, and as a result, we select this configuration. The validation values for implementing batch normalization or not implementing it is 0.0049 or 0.0042, respectively.

Subsequently, we will employ the optimal network model, as described above, to reconstruct our distance-redshift curve.

\subsection{Supernova Data} \label{subsec:sn data}
In order to reconstruct the distance function $d(z)$, we choose the latest combined sample of SNe Ia called Pantheon+
\citep{2022ApJ...938..113S}, which consists of 1701 light curves of 1550 SNe Ia, covering the redshift range $0.001<z<2.3$. For each SN Ia,
the distance modulus $\mu$ is related to the luminosity distance $D_L$ by
\begin{equation}
\mu(z)=5\log_{10}\left[\frac{D_{L}(z)}{\rm Mpc}\right]+25\;,
\end{equation}
and the observed distance modulus is
\begin{equation}\label{eq:mu_SN}
\mu_{\rm obs}(z)=m_{B}(z)+ \kappa \cdot X_{1}-\omega \cdot \mathcal{C}-M_{B}\;,
\end{equation}
where $m_B$ is the rest-frame \emph{B} band peak magnitude, $X_{1}$ and $\mathcal{C}$, respectively, represent
the time stretch of light curve and the SN color at maximum brightness, and $M_{B}$ is the absolute \emph{B}-band
magnitude. Through the BEAMS with Bias Corrections method \citep{2017ApJ...836...56K}, the two nuisance parameters
$\kappa$ and $\omega$ can be calibrated to zero. Then, the observed distance modulus can be simplified as
\begin{equation}
\mu_{\rm obs}(z)=m_{\rm corr}(z)-M_B\;,
\end{equation}
where $m_{\rm corr}$ is the corrected apparent magnitude. The absolute magnitude $M_{B}$ is exactly degenerate with
the Hubble constant $H_0$. Once the value of $M_{B}$ or $H_0$ is known, the luminosity distances $D_{L}(z)$ can be
obtained from SNe Ia.

\begin{figure}[t!]
\vskip-0.2in
\centering
\includegraphics[width=1.0\linewidth]{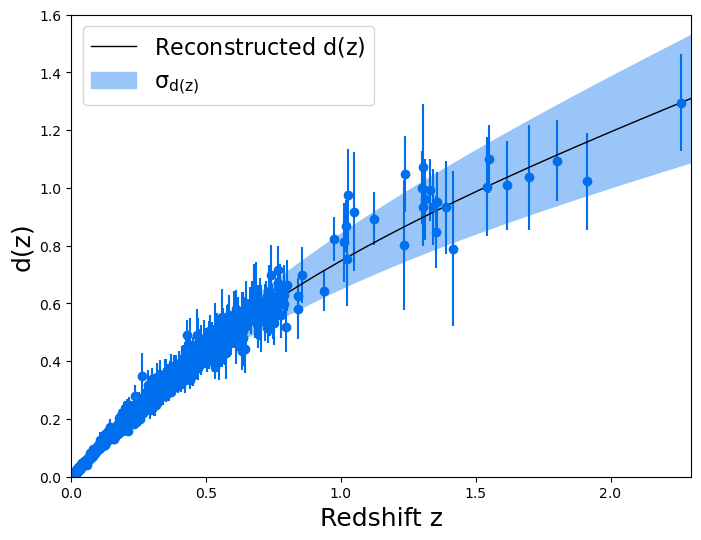}
\caption{Reconstruction of the dimensionless comoving distance $d(z)$ from Pantheon+ SNe Ia using ANN.
The shaded area is the 1$\sigma$ confidence level of the reconstruction. The blue dots with error bars
represent the observational data.
\label{rec_d}}
\end{figure}

In this work, we adopt $H_0=70\;\rm km\;s^{-1}\;Mpc^{-1}$ to normalize the SN Ia $D_{L}(z)$ data as the observational
$d(z)$. That is, $d(z)=(H_{0}/c)D_{L}(z)/(1+z)$. Note that the choice of $H_0$ has no impact on our results, since
the required distance ratio $D_{s}/D_{ls}$ (see Equation~(\ref{13})) is completely independent of $H_0$. Having obtained
the dataset of $d(z)$, we adopt ANN to reconstruct the distance function $d(z)$, and the results are shown in Figure~\ref{rec_d}.
The black line represents the reconstructed function of $d(z)$, and the shaded region is the corresponding $1\sigma$
confidence level.

\subsection{Strong-lensing Data} \label{subsec:sgl data}
Recently, Ref.~\cite{10.1093/mnras/stz1902} compiled a galaxy-scale SGL sample including 161 systems with stellar
velocity dispersion measurements, which is assembled with strict selection criteria to meet the assumption
of spherical symmetry on the lens mass model. The observational information for each SGL system are listed in
Appendix of Ref.~\cite{10.1093/mnras/stz1902}, including the lens redshift $z_l$, the source redshift $z_s$,
the Einstein angle $\theta_E$, the central velocity dispersion of the lensing galaxy $\sigma_{\rm ap}$,
the spectroscopic aperture angular radius $\theta_{\rm ap}$, and the half-light angular radius of the lensing
galaxy $\theta_{\rm eff}$.

By fitting the two-dimensional power-law luminosity profile convolved with the instrumental point spread function
to the high-resolution Hubble Space Telescope imaging data over a circle of radius $\theta_{\rm eff}/2$ centered
on the lensing galaxies, Ref.~\cite{10.1093/mnras/stz1902} measured the slops of the luminosity density profile $\delta$
for the 130 lensing galaxies in the full sample. They showed that $\delta$ should be treated as an observable for each lens in order to get an unbiased estimate of the cosmological parameter $\Omega_{\rm m}$.
Therefore, the SGL sample we adopt here is the truncated sample of 130 SGL systems with $\delta$ measurements,
for which the redshift ranges of lenses and sources are $0.0624\leq z_{l} \leq0.7224$ and $0.1970\leq z_{s} \leq2.8324$,
respectively. In this work, we use the reconstructed distance function $d(z)$ from Pantheon+ SNe Ia to calibrate
the distances $d_l$ and $d_s$ of the SGL systems. However, the SN Ia catalog extends only to $z=2.3$. As such,
we shall employ only a sub-set of the SGL sample that overlaps with the SN Ia data for the calibration. Thus, only 120 SGL systems with $z_{s}\leq2.3$ are available in our analysis.

\subsection{The Likelihood Function} \label{subsec:LF}
By using the Python Markov Chain Monte Carlo module EMCEE \citep{2013PASP..125..306F} to maximize the likelihood function $\mathcal{L}$,
we simultaneously place limits on the PPN parameter $\gamma_{\rm PPN}$, the curvature parameter $\Omega_k$, and
the lens model parameters ($\alpha_0$, $\alpha_z$, and $\alpha_s$). The likelihood function is defined as
\begin{equation}\label{19}
\mathcal{L}=\prod_{i=1}^{120}\frac{1}{\sqrt{2\pi}\Delta \sigma_{0,i}^{\rm
tot}}\exp \left [ -\frac{1}{2}\left ( \frac{\sigma_{0,i}^{\rm th}-\sigma_{0,i}^{\rm obs}}{\Delta \sigma_{0,i}^{\rm tot}} \right )^2 \right ]\;,
\end{equation}
where the variance
\begin{equation}\label{20}
\left ( \Delta\sigma_0^{\rm tot} \right )^2=\left ( \Delta \sigma_{0}^{\rm SGL} \right )^2+\left ( \Delta \sigma_{0}^{\rm SN}\right )^2
\end{equation}
is given in terms of the total uncertainty $\Delta \sigma_{0}^{\rm SGL}$ derived from the SGL
observation (Equation~(\ref{15})) and the propagated uncertainty $\Delta \sigma_{0}^{\rm SN}$ derived from the distance calibration by SNe Ia. With Equation (\ref{13}), the propagated uncertainty $\Delta \sigma_{0}^{\rm SN}$ can be estimated as
\begin{equation}\label{21}
\Delta \sigma_{0}^{\rm SN}=\sigma_0^{\rm th}\frac{\Delta D_r}{2D_r}\;,
\end{equation}
where $D_r$ is a convenient notation for the distance ratio in
Equation (\ref{13}), i.e., $D_r\equiv D_{s}/D_{ls}=d_{s}/d_{ls}$, and its uncertainty is $\Delta D_r$. With the reconstructed distance function $d(z)$,
as well as its $1\sigma$ uncertainty $\Delta d(z)$, from the SN Ia data, we can calibrate the distances ($d_{l}$ and $d_{s}$) and their corresponding uncertainties ($\Delta d_l$ and $\Delta d_s$) for each SGL system. Thus, the uncertainty $\Delta D_r$ of the distance ratio can be easily derived from
Equation (\ref{18}), i.e.,
\begin{equation}\label{22}
\begin{aligned}
\left ( \Delta D_r \right )^2= & D_r^4  \left ( \frac{\Omega_kd_l}{\sqrt{1+\Omega_kd_l^2}}-\frac{\sqrt{1+\Omega_kd_s^2}}{d_s} \right )^2\left ( \Delta d_l \right )^2 \\
& + D_r^4 \left ( \frac{d_l}{d_s^2\sqrt{1+\Omega_kd_s^2}}\right )^2\left ( \Delta d_s \right )^2  \;.
\end{aligned}
\end{equation}

\section{Results} \label{sec:result}

\begin{figure*}[ht!]
\centering
\includegraphics[width=0.8\linewidth]{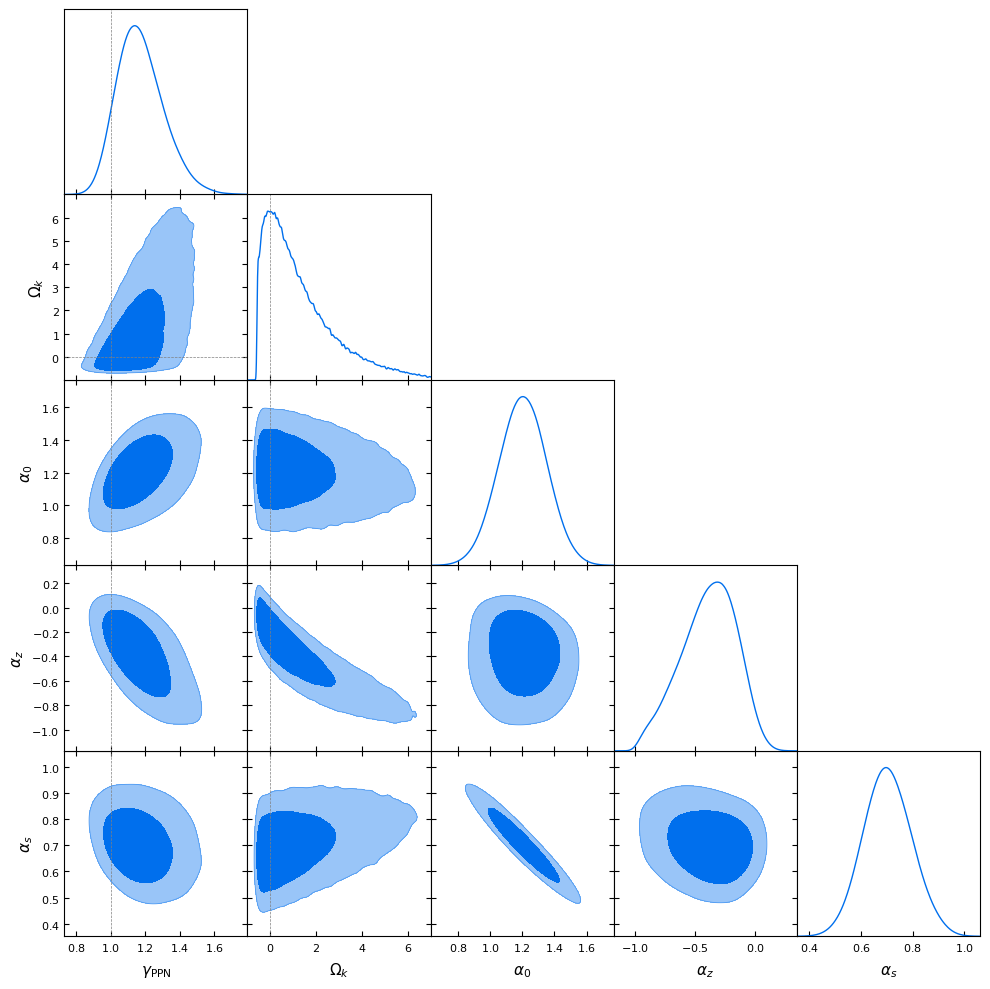}
\caption{1D marginalized probability distributions and 2D $1-2\sigma$ confidence contours for the PPN parameter $\gamma_{\rm PPN}$, the cosmic curvature $\Omega_k$, and the lens model parameters ($\alpha_0$, $\alpha_z$, and $\alpha_s$).
The dashed lines represent $\gamma_{\rm PPN}=1$ and $\Omega_k=0$, corresponding to a flat universe with the validity of GR.
\label{gamma-omegak}}
\end{figure*}

The 1D marginalized probability distributions and 2D plots of the $1-2\sigma$
confidence regions for the PPN parameter $\gamma_{\rm PPN}$, the cosmic curvature
$\Omega_k$, and the lens model parameters ($\alpha_0$, $\alpha_z$, and $\alpha_s$),
constrained by 120 SGL systems, are presented in Figure \ref{gamma-omegak}, and
the best-fitting results are listed in Table~\ref{tab:1}. These contours show that
at the $1\sigma$ confidence level, the inferred parameter values are
$\gamma_{\rm PPN}=1.16_{-0.12}^{+0.15}$, $\Omega_{k}=0.89_{-1.00}^{+1.97}$,
$\alpha_0=1.2_{-0.15}^{+0.15}$, $\alpha_z=-0.37_{-0.26}^{+0.22}$, and
$\alpha_s=0.70_{-0.09}^{+0.10}$. We find that the measured $\gamma_{\rm PPN}$
is consistent with the prediction of $\gamma_{\rm PPN}=1$ from GR, and its
constraint accuracy is about 11.6\%. While $\Omega_k$ is weakly constrained,
it is still compatible with zero spatial curvature within $1\sigma$ confidence
level. We also find that the inferred $\alpha_z$ and $\alpha_s$ separately
deviate from zero at $\sim2\sigma$ and $\sim8\sigma$ levels, confirming previous
finding that the total mass density slope $\alpha$ strongly depends on both
the lens redshift and the surface mass density \citep{10.1093/mnras/stz1902}.

\begin{table*}[ht!]
\centering
\caption{Constraint results for All Parameters with Different Priors \label{tab:1}}
\setlength{\tabcolsep}{20pt}{
\begin{tabular}{lccccc}
\hline\hline
Priors & $\gamma_{\rm PPN}$& $\Omega_k$ & $\alpha_0$ & $\alpha_z$ & $\alpha_s$\\
\hline
None & $1.16_{-0.12}^{+0.15}$ & $0.89_{-1.00}^{+1.97}$ & $1.20_{-0.15}^{+0.15}$ & $-0.37_{-0.26}^{+0.22}$ & $0.70_{-0.09}^{+0.10}$\\
$\Omega_k=0$ & $1.09_{-0.10}^{+0.11}$ &  & $1.22_{-0.14}^{+0.14}$ & $-0.20_{-0.11}^{+0.11}$ & $0.67_{-0.09}^{+0.09}$\\
$\gamma_{\rm PPN}=1$ &  & $0.12_{-0.47}^{+0.78}$ & $1.10_{-0.12}^{+0.11}$ & $-0.20_{-0.16}^{+0.15}$ & $0.74_{-0.08}^{+0.08}$\\
\hline
\end{tabular}}
\end{table*}

\begin{figure*}[ht!]
\centering
\includegraphics[width=0.6\linewidth]{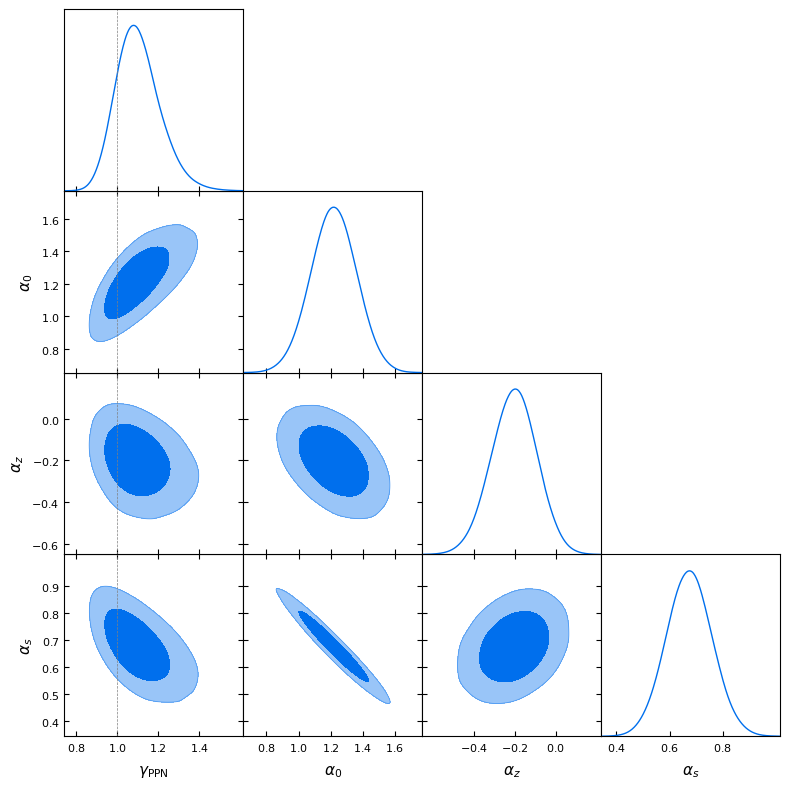}
\caption{Same as Figure~\ref{gamma-omegak}, except now for the scenario with
a prior of $\Omega_k=0$. The dashed line represents $\gamma_{\rm PPN}=1$
predicted by GR.
\label{omegak_0}}
\end{figure*}

\begin{figure*}[ht!]
\centering
\includegraphics[width=0.6\linewidth]{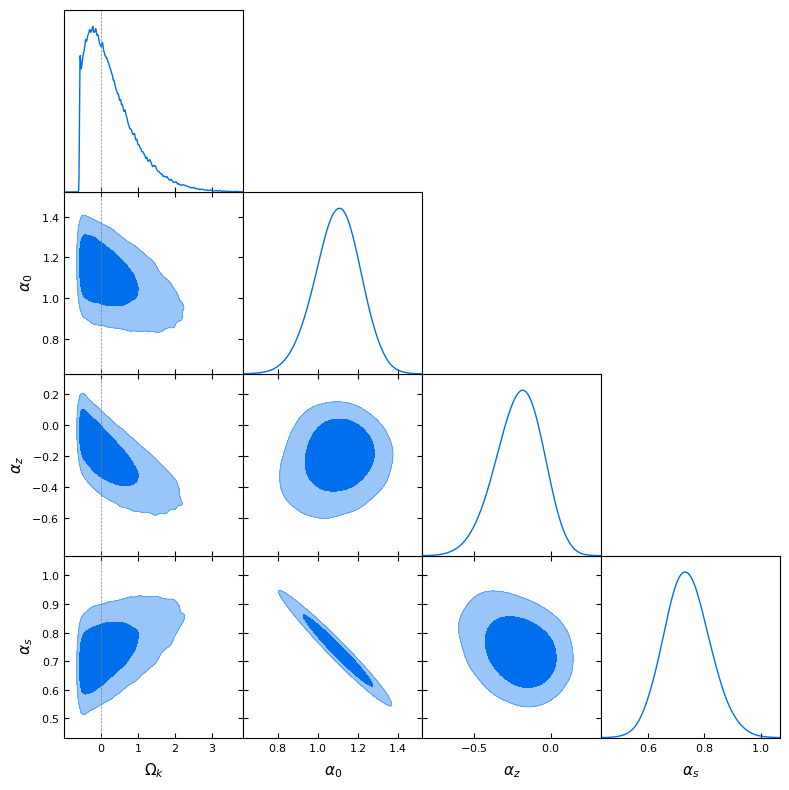}
\caption{Same as Figure~\ref{gamma-omegak}, except now for the scenario with a prior of $\gamma_{\rm PPN}=1$. The dashed line represents a spatially flat universe.
\label{gamma_1}}
\end{figure*}

We further explore the scenario of adopting a prior of flatness, i.e., $\Omega_k=0$.
For this scenario, as shown in Figure~\ref{omegak_0} and Table~\ref{tab:1},
the marginalized distribution gives $\gamma_{\rm PPN}=1.09_{-0.10}^{+0.11}$,
representing a precision of 9.6\%, in good agreement with the prediction of GR.
If instead we adopt a prior of $\gamma_{\rm PPN}=1$ (i.e., assuming GR holds)
and allow $\Omega_k$ to be a free parameter, the resulting constraints on
$\Omega_k$ and the lens model parameters are displayed in Figure~\ref{gamma_1}
and Table~\ref{tab:1}. The marginalized $\Omega_k$ constraint is
$\Omega_k=0.12_{-0.47}^{+0.78}$, consistent with a spatially flat universe.
The comparison among lines 1--3 of Table~\ref{tab:1} suggests that different
choices of priors have little effect on the lens model parameters ($\alpha_0$,
$\alpha_z$, and $\alpha_s$).

From Figures~\ref{gamma-omegak}--\ref{gamma_1}, we can see that
there is a strongly squeezed relation between $\alpha_0$ and $\alpha_s$. This
seems to indicate that to some extent one can replace $\alpha_s$ by mass density
independent $\alpha_0$, but not entirely. This may also imply that the adopted
parametrization for the total mass density slope $\alpha$ (Equation (\ref{5}))
is not the most compatible lens model. Nevertheless, it is worth pointing out
that besides the dependence of $\alpha$ on lens redshift and surface mass density,
other unexplored but important dependencies may also be found in future that can
result in a more accurate phenomenological model for lensing galaxies \citep{10.1093/mnras/stz1902},
thereby providing more reliable constraints on the PPN parameter $\gamma_{\rm PPN}$.

\section{Conclusion and Discussions} \label{sec:c&d}
Galaxy-scale SGL systems, combined with stellar velocity dispersion measurements of lensing galaxies, provide a
powerful probe to test the validity of GR by constraining the PPN parameter $\gamma_{\rm PPN}$ on kiloparsec scales.
Testing GR in this manner, however, it is necessary to know the angular diameter distances between
the observer, lens, and source. Conventionally, the required distances are calculated within the standard $\Lambda$CDM
cosmological model. Such distance calculations would involve a circularity problem in testing GR, since $\Lambda$CDM
itself is established on the framework of GR. In this paper, in order to address the circularity problem, we have
employed the DSR in the FLRW metric to estimate not only $\gamma_{\rm PPN}$ but also the spatial curvature $\Omega_k$
independently of any specific cosmological model. To calibrate the distances of the SGL systems, we have introduced a new nonparametric approach for reconstructing the distance--redshift relation from the Pantheon+ SN Ia sample using
an ANN, which has no assumptions about the observational data and is a completely data-driven approach.

By combining 120 well-selected SGL systems with the reconstructed distance function from 1701 data points of SNe Ia,
we have obtained simultaneous estimates of $\gamma_{\rm PPN}$ and $\Omega_k$  without any specific
assumptions about the contents of the universe or the theory of gravity. Our results show that
$\gamma_{\rm PPN}=1.16_{-0.12}^{+0.15}$ and $\Omega_{k}=0.89_{-1.00}^{+1.97}$. The measured $\gamma_{\rm PPN}$ is in
good agreement with the prediction of GR with 11.6\% accuracy. If we use flatness as a prior (i.e., $\Omega_k=0$),
we infer that $\gamma_{\rm PPN}=1.09_{-0.10}^{+0.11}$, representing a precision of 9.6\%. If we instead assume the conservation of GR (i.e., $\gamma_{\rm PPN}=1$) and allow $\Omega_k$ to be a free parameter, we find
$\Omega_k=0.12_{-0.47}^{+0.78}$. The measured $\Omega_k$ is consistent with zero spatial curvature, suggesting that there is no significant deviation from a flat universe.

In the literature, Ref.~\cite{2006PhRvD..74f1501B} used 15 SLACS lenses to give a result of
$\gamma_{\rm PPN}=0.98\pm 0.07$, based on priors on galaxy structure from local observations and the assumption
of a $\Lambda$CDM model with density parameters ($\Omega_{\rm m}$, $\Omega_\Lambda$)=(0.3, 0.7). Utilizing a sample of 80 SGL systems, Ref.~\cite{2017ApJ...835...92C} obtained the constraint accuracy of the PPN parameter $\gamma_{\rm PPN}$ to be 25\% under the assumption of $\Lambda$CDM with parameters taken from Planck observations. Within the same context of $\Lambda$CDM, Ref.~\cite{Collett2018Sci...360.1342C} concluded that $\gamma_{\rm PPN}=0.97\pm 0.09$ (representing a precision of 9.3\%) by analyzing the nearby lens ESO 325-G004. Through the reanalysis of four time-delay lenses,
Ref.~\cite{Yang:2020eoh} obtained simultaneous constraints of $\gamma_{\rm PPN}$ and the Hubble constant $H_0$
for flat $\Lambda$CDM, yielding $\gamma_{\rm PPN}=0.87_{-0.17}^{+0.19}$ (representing a precision of 21\%)
and $H_0=73.65_{-2.26}^{+1.95}\;\rm km\;s^{-1}\;Mpc^{-1}$. Within a flat FLRW metric, Ref.~\cite{2022ApJ...927...28L}
used 120 lenses to achieve a model-independent estimate of $\gamma_{\rm PPN}=1.065_{-0.074}^{+0.064}$
(representing a precision of 6.5\%) by employing the GP method to reconstruct the SN distances.
As a further refinement, Ref.~\cite{2022ApJ...927L...1W} removed the flatness assumption and implemented the DSR
to obtain model-independent constraints of $\gamma_{\rm PPN}=1.11_{-0.09}^{+0.11}$ (representing a precision of 9.0\%)
and $\Omega_k=0.48_{-0.71}^{+1.09}$. Note that in Ref.~\cite{2022ApJ...927L...1W} the distances of the SGL systems
were determined by fitting a third-order polynomial to the SN Ia data. Unlike the polynomial fit that rely on
the assumed parameterization, the ANN used in this work is a completely data-driven approach that could reconstruct
a function from various data without assuming a parameterization of the function. Moreover, unlike the GP method
that rely on the assumption of Gaussian distributions for the observational random variables, the ANN method has no assumptions about the data. More importantly, compared to previous results, our work yielded comparable resulting
constraints on $\gamma_{\rm PPN}$, which indicates the effectiveness of data-driven modeling based on the ANN.

Note that the dataset used here is roughly ten times larger than that used by Ref.~\cite{2006PhRvD..74f1501B}.
Yet the results obtained are almost the same level of accuracy. We attribute the absence of significantly enhanced accuracy to three primary reasons. (i) Ref.~\cite{2006PhRvD..74f1501B} modeled the probability distribution of the total mass density slope $\alpha$
as a Gaussian prior of $\alpha=1.93\pm0.08$. But, Ref.~\cite{2017ApJ...835...92C} highlighted that the most substantial source of
systematic error on $\gamma_{\rm PPN}$ is the scatter of $\alpha$. That is, without the Gaussian prior on $\alpha$, the systematic
error on $\gamma_{\rm PPN}$ would be much larger. Our worse $\gamma_{\rm PPN}$ accuracy appears to be due to the fact that
Ref.~\cite{2006PhRvD..74f1501B} set a very narrow Gaussian prior on $\alpha$, while we consider the dependence of $\alpha$ on
lens redshift and surface mass density (see Equation (\ref{5})) and adopt wide flat priors on the free parameters $\alpha_0$,
$\alpha_z$ and $\alpha_s$. (ii) Compared to our sample \citep{2006ApJ...638..703B,2006ApJ...640..662T,2006ApJ...649..599K},
the sample employed in Ref.~\cite{2006PhRvD..74f1501B} exhibits notably higher accuracy in velocity dispersion measurements.
(iii) Ref.~\cite{2006PhRvD..74f1501B} assumed $\Lambda$CDM with specific cosmological parameters to compute distances $D_s$
and $D_{ls}$, thereby lacking the uncertainty derived from distance calibration.

Looking forward, the forthcoming Large Synoptic Survey Telescope (LSST) survey, with its excellent operation performance,
holds great promise for detecting a large number of lenses, potentially reaching up to 120,000 in the most optimistic
scenario \citep{Collett2015ApJ...811...20C}. By setting a prior on the curvature parameter $-0.007<\Omega_k<0.006$,
Ref.~\cite{2017ApJ...835...92C} showed that 53,000 simulated LSST strong lensing data would set a stringent
constraint of $\gamma_{\rm PPN}=1.000_{-0.0011}^{+0.0009}$, reaching a precision of $10^{-3}\sim 10^{-4}$.
Similarly, Ref.~\cite{Lian2022ApJ...941...16L} performed a robust extragalactic test of GR using a well-defined sample
of 5,000 simulated strong lenses from LSST, yielding an accuracy of 0.5\%. In brief, much more severe constraints on
both $\gamma_{\rm PPN}$ and $\Omega_k$, as discussed in this work, can be expected with the help of future lens surveys.

\begin{acknowledgments}
We are grateful to the anonymous referee for helpful comments.
This work is partially supported by the National Natural Science Foundation of China
(grant Nos. 12373053 and 12041306), the Key Research Program of Frontier Sciences (grant No. ZDBS-LY-7014)
of Chinese Academy of Sciences, the Natural Science Foundation of Jiangsu Province (grant No. BK20221562),
and the Young Elite Scientists Sponsorship Program of Jiangsu Association for Science and Technology.
\end{acknowledgments}


\bibliography{ms}

\end{document}